\documentclass[aps,pre,reprint,groupedaddress]{revtex4-1}
\usepackage[T1]{fontenc}
\usepackage[latin9]{inputenc}
\usepackage[usenames,dvipsnames]{color}
\usepackage[english]{babel}
\usepackage{verbatim}
\usepackage{graphicx}
\definecolor{green}{rgb}{0,0.8,0.4}
\usepackage[unicode=true,pdfusetitle,
 bookmarks=true,bookmarksnumbered=false,bookmarksopen=false,
 breaklinks=false,pdfborder={0 0 1},backref=false,colorlinks=false]
 {hyperref}

\makeatletter

\usepackage{amstext}\@ifundefined{definecolor}
 {\usepackage{color}}{}

\newcommand{\ee}{\end{equation}}
\newcommand{\be}{\begin{equation}}
\newcommand{\eea}{\end{eqnarray}}
\newcommand{\bea}{\begin{eqnarray}}
\newcommand{\ea}{\end{array}}
\newcommand{\ba}{\begin{array}}
\newcommand{\bn}{\begin{note}}
\newcommand{\en}{\end{note}}
\newcommand{\bc}{\begin{center}}
\newcommand{\ec}{\end{center}}

\usepackage{graphics}\@ifundefined{definecolor}
 {\usepackage{color}}{}
\usepackage{shortvrb}
\makeatother

\begin{document}

\title{Topological determinants of self-sustained activity in a simple model of excitable dynamics on graphs}
%\titlerunning{Topological determinants of self-sustained activity}
\author{C. Fretter}
\affiliation{ 
	School of Engineering and Science,
	 Jacobs University Bremen, Germany \\
	 and \\
	 Department of Computational Neuroscience, \\
	 Universit\"atsklinikum Hamburg-Eppendorf, 
	 Hamburg, Germany} 
\author{A. Lesne}
\affiliation{ 	 
 LPTMC,CNRS, UMR 7600, \\
Universit\'e Pierre et Marie Curie, Sorbonne Universit\'es, \\4 place Jussieu, F-75005, Paris, France \\
 and \\
 GMM, CNRS UMR 5535, \\
 Universit\'e de Montpellier,\\
 1919 route de Mende, F-34294, Montpellier, France}

\author{C. C. Hilgetag}
\affiliation{ 	
Department of Computational Neuroscience, \\
	 Universit\"atsklinikum Hamburg-Eppendorf, 
	 Hamburg, Germany\\
	 and\\
Department of Health Sciences, 
Boston University, \\
Boston, USA}	 

\author{M.-Th. H\"utt}
\affiliation{ 	
	 School of Engineering and Science,
	 Jacobs University Bremen, Germany \\
	 }

\date{\today}

\begin{abstract}
Models of simple excitable dynamics on graphs are an efficient framework
for studying the interplay between network topology and dynamics.
This subject is a topic of practical relevance to diverse fields,
ranging from neuroscience to engineering. Here we analyze how a single
excitation propagates through a random network as a function of the
excitation threshold, that is, the relative amount of activity in the
neighborhood required for an excitation of a node. Using numerical
simulations and analytical considerations, we can understand the onset
of sustained activity as an interplay between topological cycle statistics
and path statistics. Our findings are interpreted in the context of
the theory of network reverberations in neural systems, which is a
question of long-standing interest in computational neuroscience.
\end{abstract}

\maketitle
\section{Introduction}
The diverse ways, in which architectural features of neural networks
can facilitate sustained excitable dynamics, is a topic of interest
both in the theory of complex networks and in computational neuroscience.
Minimal mathematical models can help to understand the generic features
of how such dynamics organize on graphs. Here we discuss a simple
numerical experiment, where we insert a single excitation into a graph
and allow it to propagate with a neuron-like discrete, relative-threshold
excitable dynamics. This numerical experiment can be seen as an \emph{in
vitro} setup of signal propagation and amplification. In particular,
it serves as a strategy for probing the mechanisms controlling the
onset of self-sustained activity in neuronal dynamics. The existence of stable regimes of sustained network activation is an essential requirement for the representation of functional patterns in complex neural networks, such as the mammalian cerebral cortex. In particular,   initial network activations should result in neuronal activation patterns that neither die out too quickly nor rapidly engage the entire network. Without this feature, activation patterns would not be stable, or would lead to a pathological excitation of the whole brain. 
Narrowing down the complex interplay of topology and dynamics to a minimal model
scenario allows us to understand microscopically, and to some extent
also analytically, the emergence of long transients and, subsequently,
self-sustained activity in networks.

In the present study, several topological determinants of sustained
activity are characterized and their range of application is delineated.
We introduce the concept of barriers, which are topological features possibly disrupting excitation propagation.
We analyze the contribution
of topological cycles, disrupting the layer-wise excitation fronts,
as soon as  excitation `holes' (corresponding to high-degree nodes that have not
reached the excitation level) start appearing in such fronts. In particular,
we propose a mechanistic understanding of some features of the response
curve (i.e. the number of successful excitation propagation events
as a function of the relative threshold $\kappa$): transition values
of $\kappa$ and levels reached by the successful propagation events.
This in turn allows a quantitative prediction of these features from
the detailed knowledge of the network topology.

The transient sustained activity seen in our excitable model is reminiscent
of a biological phenomenon termed \emph{network reverberation}, that
is, the temporarily sustained activity induced by a specific
stimulation of a neural circuit. The concept is related to the concept
of \emph{neural assemblies} introduced by Hebb \cite{hebb1949}. The
intuitive application of such reverberations lies in dynamic
memory circuits, that is, online (working) memory based on dynamic
patterns, rather than long-term memory that may be encoded in the
synaptic weight distribution of the network. Indeed, one can see transiently
sustained activity in specific cortical regions (e.g., prefrontal
and posterior parietal cortex) related to working memory tasks, such
as a delayed matching-to-sample task. The predominant idea is that
reverberations are expressed as \emph{dynamic attractors} of transiently
stable increased activity, particularly due to locally increased synaptic
strength \cite{Wang:2001wk}. This idea also provides a link between the dynamic patterns encoding short-term (working) memory and the synaptic weight changes underlying long-term memory.
However, there exists an extensive debate
on the specific circuitry and parameters underlying the reverberations,
e.g. \cite{HadipourNiktarash:2003gx,Muresan:2007ez,Tegner:2002bc}.

Using discrete dynamical models to explore relationships between network
architecture and dynamics has provided some key insights into
the functions of complex networks in the past, e.g. Boolean models
for gene regulatory networks \cite{bornholdt2005systems} and SIR and
SIS models for epidemic diseases in social networks \cite{pastor2001epidemic}. 

How network topology can facilitate the self-sustainment of excitable dynamics on graphs is a fundamental question about the organization of dynamics on graphs \cite{roxin04,Deco:2009p6486,Deco:2011p775,ptrs2014}. 
The role of cycles in excitable dynamics on graphs has received a remarkable amount of attention in the last years \cite{Qian:2010a,Mcgraw:2011jr,liao:2011}, in particular in Computational Neuroscience \cite{lewis2000,vladimirov}. 
Cycles have been implicated in maintaining activity in a network \cite{vladimirov,Mcgraw:2011jr}. In \cite{Garcia:2012ey,Qian:2010b} this role of cycles in graphs has been compared to spiral waves in spatiotemporal pattern formation (see also \cite{ptrs2014}), as the cycle length (similarly to the size of the spiral core) needs to match the refractory period of the excitable units. 
In \cite{garcia2014} it has furthermore been shown that, counterintuitively, the successful usage of \textit{long} cycles contributes to the sustainment of activity in a graph.

Here our focus is not on cycle usage, but rather on the initial perturbations of the coherent propagating wave front that subsequently leads to the activation of cycles and the onset of sustained activity. 
Again resorting to the analogy to spatiotemporal pattern formation, we are here exploring the transition from target waves to spiral waves triggered by some heterogeneity in the system. In our case, the source of this heterogeneity is complex network topology.

The respective influence  of hubs (high-degree nodes) and modules in shaping activation patterns
has been investigated with a focus  on the discriminating  interplay with spontaneous excitations \cite{MullerLinow:2008ia,hutt2009interplay}. 
The role of cycles in storing excitations and favoring self-sustained activity has been yet elucidated only in a deterministic model of excitable neural networks \cite{Garcia:2012ey}. 
A phenomenon of stochastic resonance (noise-facilitated signal propagation)  has been evidenced in  so-called `sub-threshold' networks, that is, for which a single input excitation does not propagate to the output nodes \cite{hutt2012stochastic}. However,  knowing that there a limit to propagation
at some transition value  of the  parameter $\kappa$  of the excitable dynamics, henceforth denoted $\kappa_c$, is not sufficient: it is now necessary  to understand  what controls the onset or failure of excitation propagation and how the network itself produces such a threshold behavior.

Our principal goal is to gain a mechanistic understanding of the main dynamical processes underlying the two thresholds. For this investigation we employ a discrete-time three-state model of excitable dynamics already used in \cite{MullerLinow:2006ex,MullerLinow:2008ia,hutt2009interplay,hutt2012stochastic} for analyzing the relationship between network topology and excitable dynamics. 

Such cellular automata on graphs are a well established method for probing the relationship between network architecture and dynamics (see, e.g., \cite{Li:1992td,Marr:2030p1915,marr2012}). It is clear that discrete models need to be used with a certain care, as some dynamical effects can indeed be artifacts of the (time and state) discretizations. However, the possibility to unambiguously define events like co-activation or sequential activation make such discrete models powerful tools for exploring the mechanisms by which  network architecture dictates some key features of excitable dynamics (see, e.g., \cite{Garcia:2012ey,messe2014}). 

Finding a suitable balance between realism and genericity in modeling excitable dynamics is an important question in computational neuroscience (see, e.g., \cite{Giaquinta:2000vg,Deco:2011p775,Garcia:2012ey,messe2014}. 
We expect, that the mechanisms / processes here identified (underlying the two thresholds) are elementary enough to be universal and independent of the specific model of excitable dynamics. Like in other fields (e.g. epidemic diseases, \cite{moreno02}, or gene regulation, \cite{bornholdt2005systems}), the minimal model (a three-state cellular automaton on a graph) enables us to extract a few `stylized facts' about the onset of self-sustained activity, by separating the logical organization (both, on the structural and on the functional level) from the physiological details, how this logical organization is implemented. 

Our study focuses on a discrete three-state model representing a stylized
biological neuron. While it is conceptually similar to an SIR model,
our model is different, both in its biological motivation and (due
to the SIR infection probability as the key parameter and source of stochasticity) in its dynamical
behavior. The remainder of this paper is structured as follows: First, we briefly summarize the mathematical model, as well as our prediction strategy  of the excitation propagation failure or amplification encoded in network architecture (Section \ref{methods}). In Section \ref{results} we describe the generic properties (transition values of the parameter, excitation levels) of the response curve generated by a single inserted excitation, as well as our prediction results for different network architectures. Section \ref{discussion} discusses, how these features arise from an interplay of cycles and paths statistics in networks. Lastly, we summarize the implications of our findings for computational neuroscience.

\section{Methods}
\label{methods}

\subsection{Details of the numerical experiment}

We start our numerical experiment with a single, randomly chosen
input node receiving one excitation, and then observe the propagation of excitations to an output
node, selected at random from the nodes at the largest distance from the input node. 

We use a three-state cellular automaton model of excitable dynamics. Each node can be in an susceptible/excitable  ($S$), 
active/excited ($E$) or refractory ($R$) state. 
The model operates on discrete time and employs the following synchronous update rules: For a node $i$ with $k_i$ neighbors, the transition from $S$ to $E$ occurs, when at least $\kappa k_i$ neighbors are active. The parameter $\kappa $ thus serves as a relative excitation threshold. In such a relative-threshold scenario, low-degree nodes are therefore easier to excite (requiring a smaller number of  neighboring excitations) than high-degree nodes. In neuroscience, there is some evidence that this is a plausible excitation scenario,
as neurons can readjust their excitation threshold according to the input \cite{azouz2000dynamic},  which typically leads to spike frequency adaptation \cite{benda2003universal}, and  effectively amounts to a relative input threshold. This model has also been investigated in \cite{hutt2012stochastic}.
After a time step in the state $E$ a node enters the state $R$. The transition from $R$ to $S$ occurs  stochastically with the recovery probability $p$, leading to a geometric distribution of refractory times with an average of $1/p$.. Initially all nodes are in the susceptible state $S$. The model does not allow spontaneous transitions from $S$ to $E$ (i.e., compared to previous investigations \cite{MullerLinow:2006ex,MullerLinow:2008ia,hutt2009interplay}, the probability $f$ of spontaneous excitations is set to zero). 
Therefore, the stochasticity of
the dynamics is entirely due to the stochastic recovery, controlled by the recovery probability $p$. For $p$=1, we have a deterministic model (similar to the one discussed in \cite{Garcia:2012ey}; there, however, a single neighboring excitation was sufficient to trigger transition to $E$, corresponding to $\kappa \rightarrow 0$).

Previous investigations have considered the case of an absolute threshold,  where a fixed  number (set to one in previous work) of excited neighbors triggers the activation of a susceptible node. They have shown the key role of {\it hubs}
 as organizing centers of the activity  \cite{MullerLinow:2008ia,hutt2009interplay}. In contrast, for
 a relative threshold, there is no amplification
due to a potentially increased excitability of high-degree nodes. For a given node, there
is moreover a balance between a sufficient number of excited neighbors
and the number of susceptible neighbors able to propagate the excitation.
Overall, the amplification rate at a given node is bounded by $(1-\kappa)/\kappa$. 

We recorded the accumulated excitation of one among the output nodes,
during a fixed duration $T$, and plot the result as a function of $1/\kappa$, noting that $1/\kappa$ gives the maximal degree a susceptible node
can have to be excited by a single excited neighbor. Any node of degree
higher than $1/\kappa$ appears as a {\it barrier}, that is, a node for
which having a single excited neighbor is not sufficient to get excited.

We adopt a layered view (as in \cite{hutt2012stochastic}), according to
the shortest distance of the nodes to the input node: the first layer
contains the neighbors of the input node, the second layer its second
neighbors, and the final layer all the possible output nodes. 

Hubs are more likely to be located in the first layer, because even
if the input node is not a hub, it will point to a hub with a large
probability.  Using this layered view is furthermore motivated by the fact that, due to the refractory period, excitation propagates layer-wise at low enough $\kappa$.

\subsection{Prediction strategy}

We expect that the prediction of the output signal, in such a finite
setting, is not accessible to mean-field prediction. 
This situation is not accessible to a mean-field investigation. In the past, various forms of mean-field studies have been successful in elucidating relationships between network topology and dynamics. Mean-field models have for example demonstrated a similar dynamical effect of shortcuts as of spontaneous excitations \cite{graham03}, the qualitative change of excitation density with network connectivity \cite{MullerLinow:2006ex}, and the competition of waves (centered around hubs) and modular activation (synchronous activity within modules) in hierarchical networks \cite{hutt2009interplay}. 

However, already for the phenomenon of noise amplification of excitations \cite{hutt2012stochastic}, which is related to the transitions explored here, such a mean-field approach was found incapable of reproducing the qualitative features of the dynamics (cf. Figure 6B in \cite{hutt2012stochastic}). 

This expectation
motivates us to consider single realizations of the network and investigate the excitation propagation and output signal, both
on general qualitative grounds and through quantitative simulation.
Accumulating the results obtained with several network realizations
(and for a given network, several choices of the input and output
nodes), we
then compute an average prediction quality for different classes of
networks, as a function of the network size (number of nodes $N$), network
density (number of edges $M$) or recovery probability $p$. As networks, we take Erd\H{o}s-R\'enyi (ER) graphs and Barabasi-Albert (BA) graphs (generated with preferential attachment \cite{Barabasi:1999uu}). 
In parallel,
we give general insights on the barrier pattern corresponding to a
given input node within the associated layered representation of the
network. Any qualitative variation of this pattern from node to node
in a real neural network would hint at a functional significance (and presumably
evolutionary adaptation) of this pattern. At the same time, refined (higher-order) mean field
approaches can provide us with estimates of, for instance, the importance
of multiple excitations and other dynamical features disrupting simple
predictions based on specific topological features. In particular
we will adapt mean-field models from \cite{hutt2009interplay,hutt2012stochastic}
to the present situation to estimate the dependence of such effectson network parameters.

\subsection{Definition of barriers}

During the propagation from the input node to the output node, excitations
encounter barriers, the stronger the higher the degree.  We define {\it barrier strength}
as the minimal number $n$ of active neighbors
required for its excitation. It depends on the barrier degree $k$ (the larger the higher the degree) but also on $\kappa$.
Some of the barriers might not find the required number of active neighbors in the current  dynamical states of
the network and fail to propagate excitation.
% These are nodes on the path, which this excitation
%cannot overcome in the current configuration of dynamical states in
%the network. The strongest barrier is given by the highest degree.
Determinants of successful propagation will thus involve barrier statistics
and path multiplicities.

%Barriers are a local feature. We could partition the set of nodes
%into subsets $B_{n}$ of barriers of strength $n$, that is, requiring
%at least $n$ excited neighbors to allow for the propagation of an excitation. 

%Barriers can be characterized by their strength, stepwise
%increasing with the degree, and their centrality, providing an estimate of
%their probability of being encountered in paths going from the input
%node toward the output nodes. While the former (the degree) is a purely
%local feature of the node, the latter (centrality) involves the overall
%architecture of the network.

What matters for signal propagation is not only
the strength of a barrier but also  the number $k^{in}$
of incoming links from the next upper layer, described through the
conditional probability $\rho(k^{in}|k)$ given the degree $k$ of the barrier.
The probability that a node is a barrier of strength $n$, but does
not act as  an obstacle to signal propagation, is thus
\begin{equation}
\sum_{k>(n-1)/\kappa}^{k\geq n/\kappa}\rho(k)\sum_{k_{i}n=n}^{k}\rho(k^{in}|k)\alpha(n|k^{in})
\end{equation}
where $\alpha(n|k^{in})$ is the probability to have $n$ active nodes among the $k^{in}$ neighbors of the barrier in the next upper layer. This probability can be
% of passing a  barrier of strength $n$, 
%for a node to be a barrier of strength $n$ is $\rho((n-1)/\kappa<k<n/\kappa)$,
%however this quantity is of no use, even to compute the overall average
%probability. Indeed, what matters is both the strength of the barrier,
%to know how many excitations are required, and the number $k^{in}$
%of incoming links from the next upper layer, described through the
%conditional probability $\rho(k^{in}|k)$. The computation, thus,
%yields a probability of circumventing a barrier of strength $n$ (that
%is, a probability that a node is a barrier of strength $n$, but does
%not act as such in the excitable dynamics) equal to 
%\begin{equation}
%\sum_{k>(n-1)/\kappa}^{k\geq n/\kappa}\rho(k)\sum_{k_{i}n=n}^{k}\rho(k^{in}|k)\alpha_n(k^{in})
%\end{equation}
%where $\alpha_{n}(k^{in})$ is the probability that a node of given
%number $k^{in}$ of incoming nodes from the upper layer receives at
%least $n$ excitations, 
computed in a mean-field approximation.
The argument that a node get excited if its average number $kc_{E}$ of
active neighbors is larger than $k\kappa$, leading to the mean-field evolution
equations (where $H$ is the Heaviside function): $c_E(t+1)=c_S(t)H[c_E(t)-\kappa]$, together with
 $c_{S}(t)=1-c_{E}(t)-c_{R}(t)$ and $c_{R}(t)=c_{E}(t)/p$. This yields a steady-state activity density
$c_{E}^{*}=p/(2p+1)$ (provided $\kappa<p/(2p+1)$) \cite{hutt2012stochastic}. 
It comes:
\begin{equation}
\alpha(n|k^{in})=\sum_{j=n}^{k^{in}}\left(\begin{array}{c}
k^{in}\\
j
\end{array}\right)(c_{E}^{*})^{j}(1-c_{E}^{*})^{k^{in}-j}
\end{equation}
This computation implicitly assumes that the propagation is consistent
with the layer view, moving forward in a coherent way, like a front
crossing as a whole each layer. What  then matters to get concurrent
excitations is the presence of diamond motifs along the paths. We
may alternatively consider that the excitations wander along complicated
paths. We expect that numerous excitation holes exist for $\kappa$ near $\kappa_{c}$,
which totally destroys the image of an excitation front propagating
layer-wise. In this new view, the network is well described by an
homogeneous activity density $c_{E}^{*}$, and the excitation could
reach a barrier of degree $k$ by any of the $k$ edges, not only those coming from the next upper layer. The probability
of passing a barrier of strength $n$ then writes simply $\sum_{k>(n-1)/\kappa}^{k\geq n/\kappa}\rho(k)\alpha(n|k)$
(it jointly accounts for the probability that a node is such a barrier). 
%In the model with a relative threshold, nodes such that $\bar{n}_{\kappa}\leq2$
%can only be passed if multiple excitations arrived. We consider that
%the excitations wanders along complicated paths, expecting that numerous
%holes exist for $\kappa$ near $\kappa_{c}$ which destroys the picture
%of an excitation front propagating layer-wise. In this view, the network
%is well described by a stationary mean-field excitation density $c_{E}^{*}=p/(2p+1)$,
%valid provided $\kappa<p/(2p+1)$. This value originates from the
%argument that a node get excited if its average number $kc_{E}$ of
%excited neighbors is larger than $k\kappa$, leading to the evolution
%equation (where $H$ is the Heaviside function): $c(E,t+1)=c(S,t)\left[f\;+\:(1-f)H(c(E,t)-\kappa\right]$
%This mean-field stationary solution will be used to quantitatively
%describe the surroundings of a path starting from the input node and
%the probability that multiple concurrent excitations allow excitation
%to propagate up to the output node. The excitation could reach a barrier
%of degree $k$ by any of the $k$ edges. The probability of passing
%a barrier of strength $n$ then is simply given by $\sum_{k>(n-1)/\kappa}^{k\geq n/\kappa}\rho(k)\alpha_{n}(k)$. 
This latter probability has to be summed over all barrier strengths
$n\geq2$, to get the probability that multiple concurring excitations allow signal
to propagate up to the output node:
\begin{equation}
P_{multiple}=\sum_{n\geq2}\;\;\sum_{k>(n-1)/\kappa}^{k\geq n/\kappa}\rho(k)\;\sum_{j=n}^{k}\left(\begin{array}{c}
k\\
j
\end{array}\right)(c_{E}^{*})^{j}(1-c_{E}^{*})^{k-j}
\end{equation}
This expression will be used in the following to estimate the error in one of our predictions, which relies on (topologically) estimating barrier strengths based on single excitations.

\section{Results}
\label{results}

\subsection{Generic properties of the response curve}
\begin{figure}
\includegraphics[width=0.98\columnwidth]{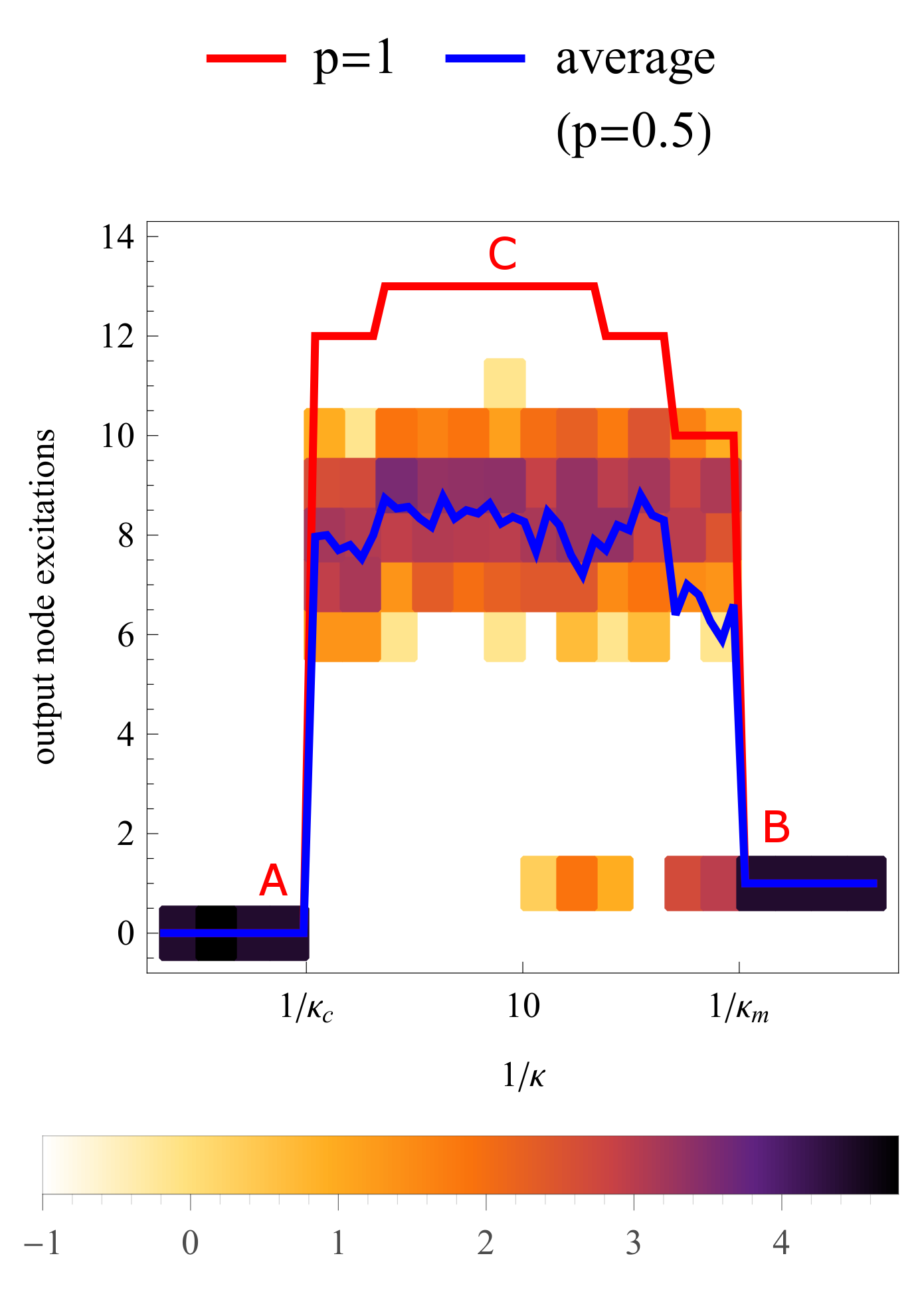} \caption{\label{fig:curve} Accumulated output excitations during a fixed duration
$T$, for a single input node and a single output node, as a function
of the inverse $1/\kappa$ of the relative threshold for $p=1$ (deterministic dynamics, red curve) and $p=0,5$  (stochastic recovery, heat map overlay of $30$ such curves and as a blue curve the average over all simulation runs entering the heat map). Initially all nodes were susceptible. Transition points
A, B and level reached, C, are delineated, and plausible explanations based on network local topology discussed in Section~\ref{sec:features}.}
\end{figure}

The system under discussion is thus an excitable network, together with the choice of an input  and an output node. The response curve of the system is the measurement of excitations at the output node as a result of a single excitation inserted at the input node.  Figure \ref{fig:curve} provides an example of such a response curve (together with a heat map overlay of many such curves), in which the generic features of these response curves are clearly visible. In the following, we will discuss the critical value $\kappa_c$ for the onset of sustained activity (point A in Figure \ref{fig:curve}), the second transition point in the threshold, $\kappa _m$, marking the boundary between the sequential excitation of layers and a turbulent self-sustained activity (point B in Figure \ref{fig:curve}), as well as the height of the response curve between these two transition points (marked as C in in Figure \ref{fig:curve}).

\begin{figure}
\includegraphics[width=0.49\columnwidth]{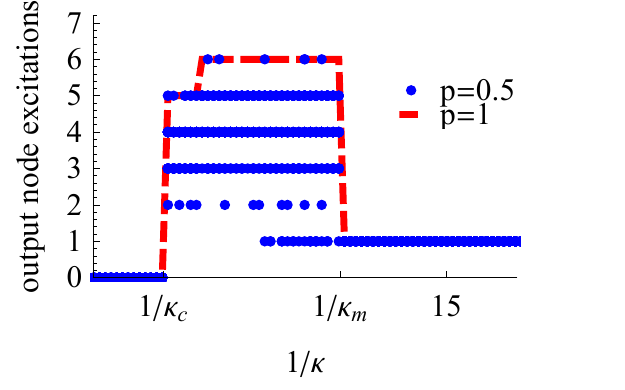}\includegraphics[width=0.49\columnwidth]{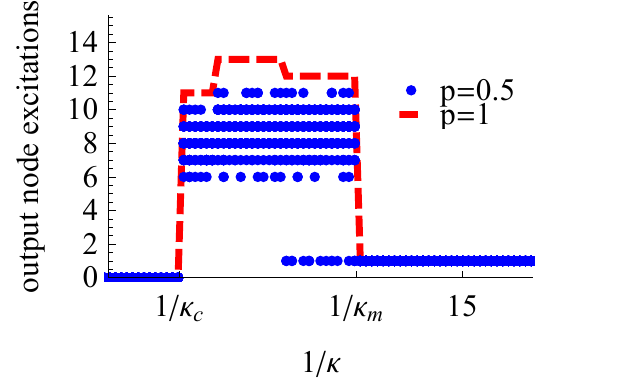}

\includegraphics[width=0.49\columnwidth]{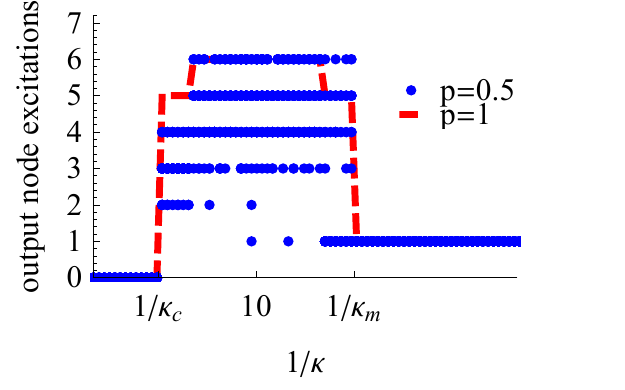}\includegraphics[width=0.49\columnwidth]{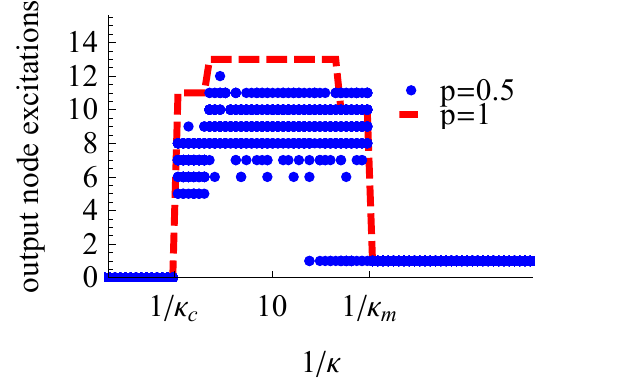}

\caption{Some examples of response curves for the stochastic  (in blue, $p=0.5$) and the deterministic case (in red, $p=1$). The graph is (ER, $M=80,$ $N=320$) for the left column and
(ER, $M=80,$ $N=800$) for the right column.}
\label{fig:curveex}
\end{figure}

The difference between the behavior observed
in case of a deterministic dynamics ($p=1$) and one where the random
recovery ($p<1$) introduces an amount of stochasticity is enlightening.
For $p<1$, one observes randomly distributed node failures as well as non zero output in the range  $\kappa_{m}<\kappa<\kappa_{c}$ (region C).
In fact the deterministic case delimits the possibility space of the
stochastic case: All excitation levels that are possible for the deterministic  dynamics
are in principle achievable in the stochastic case, if the right nodes
are susceptible again at the right moment. 
Inside of the `accessible region' situations where a higher excitation
level is achieved in the stochastic case, because a refractive node
makes a `faster' pacemaker accessible are conceivable. This exceptional
event is rarely observed experimentally.

\subsection{Prediction of response curve features}

\label{sec:features}

As pointed out in the previous Section, we can distinguish three parts or features in the curve at increasing
$1/\kappa$, denoted  A, B and C in the exemplary response curve shown in Figure~\ref{fig:curve} and in the additional examples in 
Figure \ref{fig:curveex}.
For each part of the curve, our qualitative explanation will be supported
by the comparison of some  quantitative prediction  derived from network topological features with a large
sample of simulation data (obtained with every possible input node
for at least $50$ different networks of various average degree). Comparisons are visualized as scatter
plots showing the (topologically) predicted transition value of $1/\kappa$
and the actually observed one in the simulated dynamics.
Comparisons are then be made  more quantitative by
computing prediction quality (integrated over many networks) as a function of the relevant parameters, e.g.  the average degree.

We denote $\kappa_{m}$ the value where the curve of accumulated excitations
per output node markedly departs from 1 and has its first peak (point B), and
$\kappa_{c}$ the value where the curve of accumulated excitations
per output node goes to 0 (point A). Due to the definition of the relative
threshold $\kappa$, the quantities $1/\kappa_{m}$ and $1/\kappa_{c}$ take only
integer values when the average degree of the graph is varied (by varying the edge count $M$ at fixed number of nodes $N$).

\subsubsection{\texorpdfstring{Onset of excitation propagation (transition point A, $\kappa=\kappa_c$)}{Onset of excitation propagation (transition point A, k=kc}}

All curves display a critical threshold value $\kappa_{c}$ 
%(point A in Figure~\ref{fig:curve})
 for the propagation
of a single excitation from the input node to the output nodes. 
This threshold behavior in the absence of noise is analogous to an
epidemic threshold. It does not depend on the value of $p$. 
For a finite network, the transition value $\kappa_c$ is a random variable depending on the realization
of the network, the choice of the input node and of one among the
possible output nodes, and of the initial configuration (here all
nodes are initially susceptible).

\begin{figure}
\includegraphics[width=0.98\columnwidth]{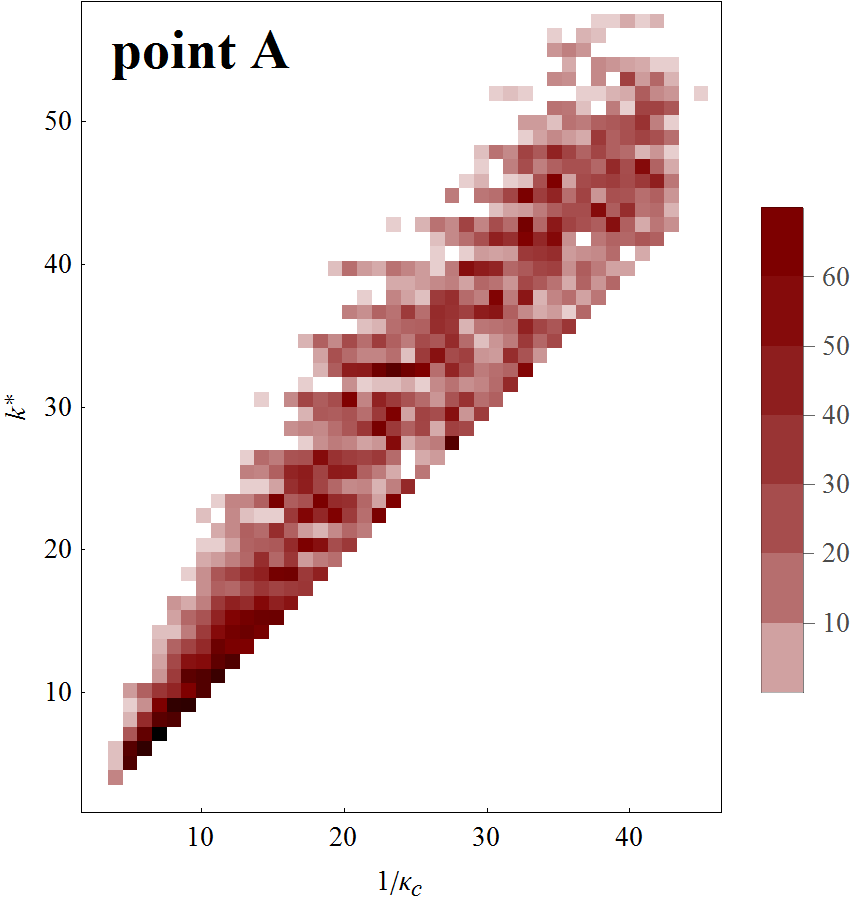}

\caption{Density histogram of the predicted $k^{*}$ for the limit of sustained
activity as a function of $1/\kappa_{c}$. The different values of $\kappa_c$ are obtained by running the dynamics on different network realizations and for different input nodes, while observing the topological observable $k^{*}$. Data are aggregated
over graphs having $N=80$ nodes and $M=100...2000$ edges. The prediction is obtained by considering that
 the node with the highest degree on the easiest path (along which the
maximal degree is minimal) from the input node to the output node
is limiting.\label{fig:histogramkcpred}}
\end{figure}

For $\kappa>\kappa_{c}$, before point A in Figure~\ref{fig:curve},
$\kappa$ is so large that each possible path from the
input node to an output node contains a barrier, that is, a node of
large degree that cannot be  excited by only one propagation excitation, 
%straightforwardly excited by the propagating signal,
 and therefore no excitation reaches the output nodes. 
The network is then termed \emph{sub-threshold} (as regards to
$1/\kappa$, which can be roughly interpreted as a transmission probability).
 In other words, a sub-threshold situation could mean that on each linear path,
there exists a node such that $k>1/\kappa$. 
Let $k^{*}$ be the largest degree encountered on the easiest paths
to the output nodes, that is, the smallest over all paths to the output
node of the maximal degree encountered along the path. Then the
onset of excitation propagation is expected to arise for a value $1/\kappa_{c}=k^{*}$.
 This prediction is tested in  Figure~\ref{fig:histogramkcpred}, where the topological observable $k^{*}$ is plotted as a function of the dynamic observable $1/\kappa_{c}$  for different networks.
Another approach to predict $\kappa_{c}$ is to apply
the same reasoning (largest degree encountered on the easiest path)
but to not consider only the output node, but all nodes on the last
layer. An improved prediction for  $1/\kappa_{c}$ could thus be
 $k^{**}$, the minimum of
the largest degrees encountered on the easiest path from the input
node to any node in the output layer.
However, the condition becomes
less stringent, if the signal propagation
activate redundant paths of the same length, so that more than one
excitation may spontaneously arrive at a given node. We thus expect that $k^{*}$ and $k^{**}$
would give only an upper bound on  $1/\kappa_{c}$, as supported by the asymmetry seen in  Figure~\ref{fig:histogramkcpred}.

Looking at the system size dependence of our prediction quality, the most interesting phenomenon is the reduction of quality for larger BA graph due to many competing hubs (see Supplementary Material). 

\begin{figure}
\includegraphics[width=0.47\textwidth]{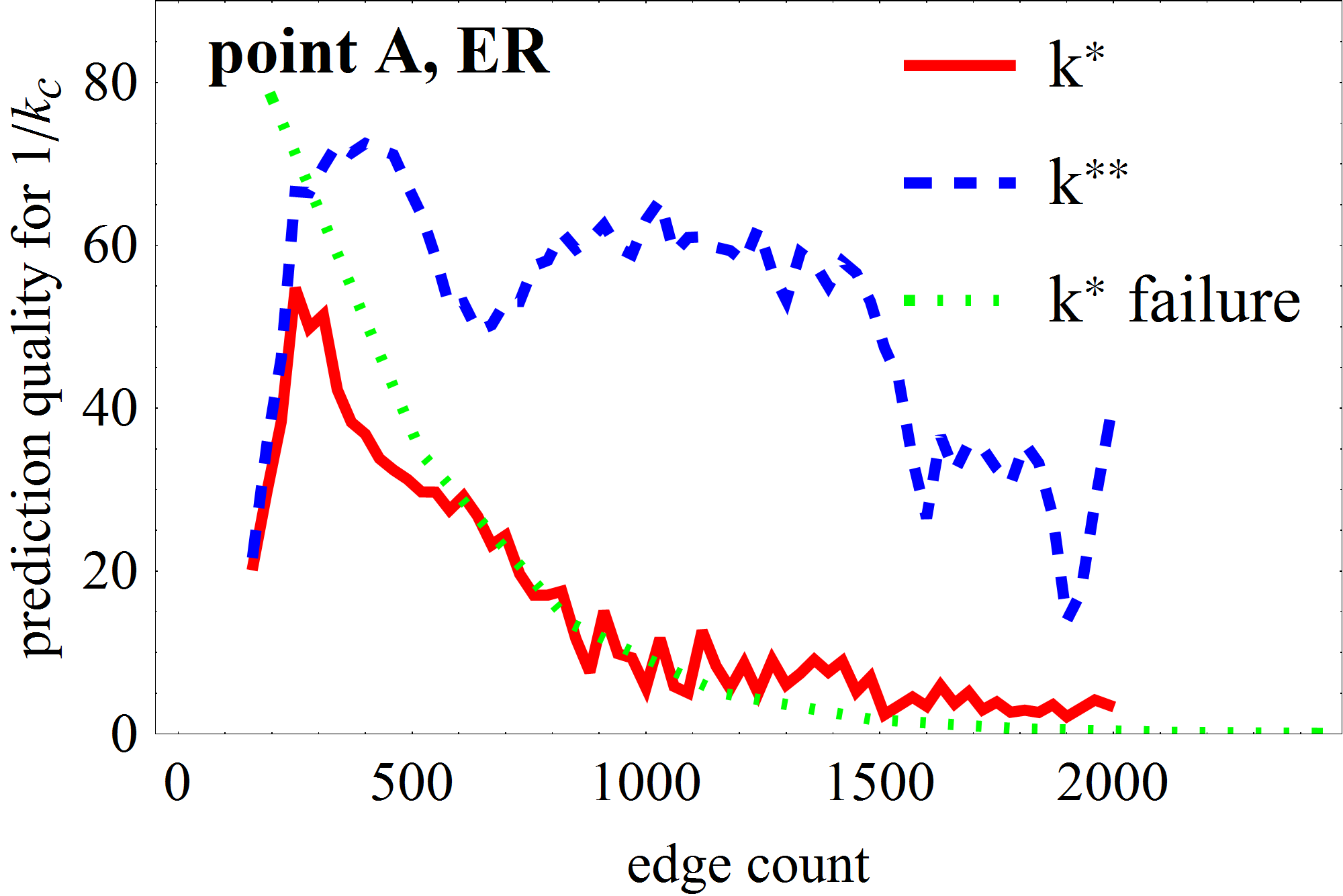}
\includegraphics[width=0.47\textwidth]{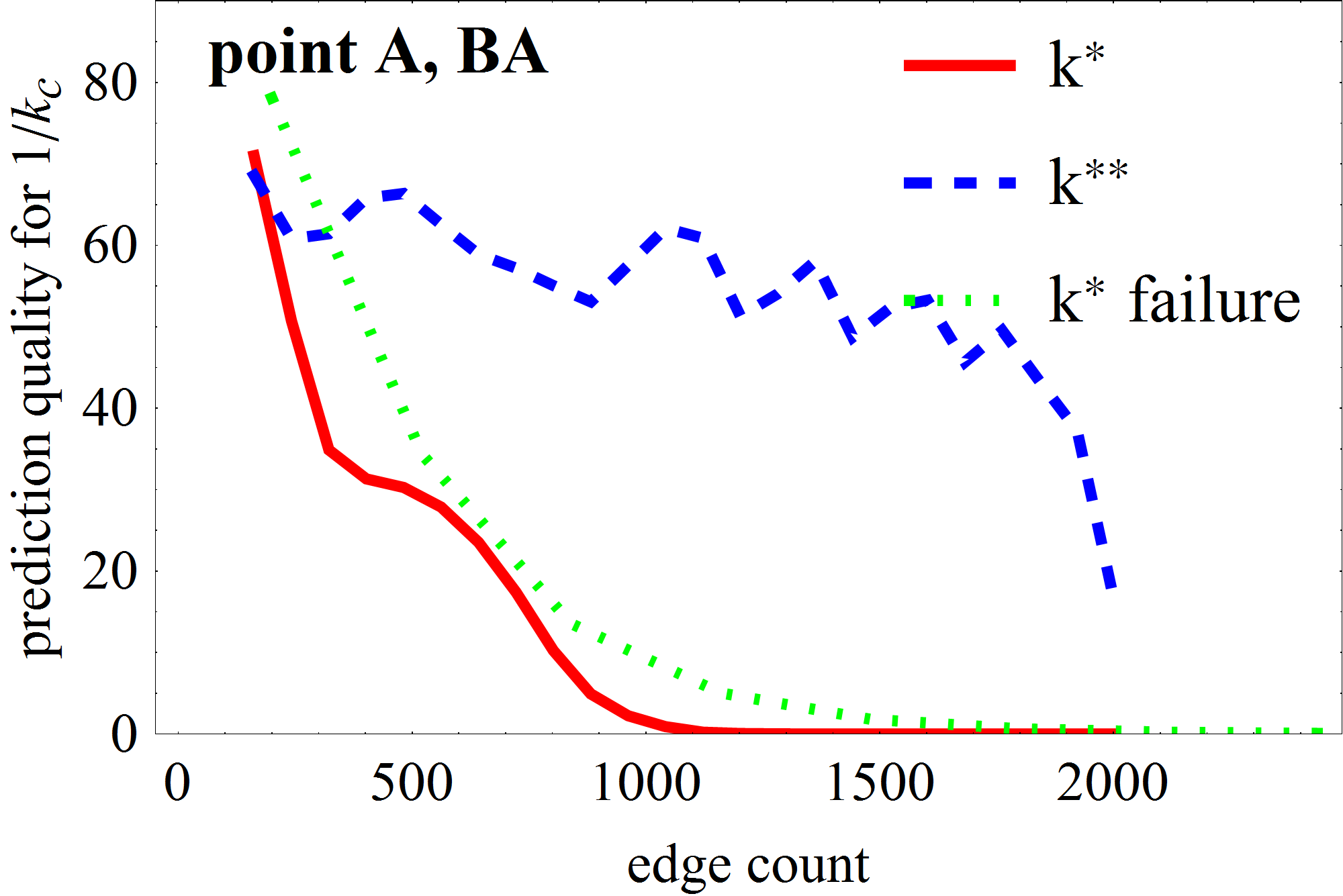}

\caption{A plot of the prediction quality for the limit $1/\kappa_c$ of sustained activity (transition point A).
The data are obtained by scanning graphs having $N=80$  nodes and $M=100...2000$ edges. We shoe the predictions $k^*$ (solid red curve) and $k^{**}$ (dashed blue curve). Additionally the dashed green line indicates the expected failure of $k^*$ due to multiple excitations. Upper Figure: ER graphs; lower Figure: BA graphs. \label{figure4}}
\end{figure}

%We here employ a mean-field computation (see Methods) to estimate
%the failures of our prediction (related to the maximal degree along
%the easiest path, that is, the strongest barrier along the easiest
%path). 
%

The prediction quality is defined as  the
percentage of cases where $\kappa_m$ or $\kappa_c$ are predicted correctly. This is determined by comparing the topological 
prediction for $1 / \kappa $ to the numerical result. 
The numerical result is obtained by a binary search in the space of $\kappa$. 
The inverse of this number is then rounded to the next integer, allowing an exact comparison. Note that, as we have observed that the transition value  $\kappa_m$ does not depend on the value of $p$, we here
use $p=1$ (the deterministic case) to make the binary search reliable.

A barrier may be passed in the case where two concurrent
excitations reach it.  The probability
of such an event, that contributes to the discrepancy between our
prediction and the observed value (and thus to the prediction quality),
cannot be computed exactly.  However,  based on Eqs (1) - (3) (see Methods), we obtain a mean-field estimate of the
importance of multiple excitations.

On this basis we can evaluate how multiple excitations contribute
to the reduction observed in the quality of the prediction $1/\kappa_{c}=k^{*}$ with increasing link density in the graphs (see Figure \ref{figure4}).

The mean-field prediction of the effect of multiple excitations qualitatively explains the decrease of the prediction. As the mean-field approach is less reliable in the regime of very low excitation densities, we here use an intermediate value of $p$ ($p$=0.5) for the mean-field prediction. Even higher values show a similarly favorable comparison with the numerical simulation.

Note that this effect, the contribution from multiple excitations, does not explain the falsely predicted cases for sparse graphs. There, the difference to $100$ percent prediction quality must be due to the more complicated layer
structure of sparse graphs. This observation is consistent with the fact that the discrepancy appears for the ER graph, but not for the BA graph, that has a more stable layer structure due to its hubs.

\subsubsection{\texorpdfstring{Transition between layer-wise propagation and sustained activity (transition point B, $\kappa=\kappa_m$)}{Transition between layer-wise propagation and sustained activity (transition point B, k=km)}}

By construction of the input-node-centered layer representation
of the network, there are no shortcuts between non adjacent layers. 
At first, the excitation injected at the input node travels layer-wise,
forming an excitation front reaching at each step a deeper layer.
A jump arises in the output signal  at some value
$1/\kappa_{m}$ (transition point B). 
Typically a high-degree node, acting as a barrier, is
not excited when the excitation front reaches its layer, and remains
susceptible, leaving a susceptible `hole' in a layer
of refractory nodes. The amplification observed at point B, in $\kappa=\kappa_m$, and explained as the appearance of the
 first cycle, is sharp.
%appearance of the first cycle corresponds to a sharp increase in the number of outgoing signal excitations. 
This means that the cycle is traveled several times, or that other cycles can be excited
after that the first one has stored excitation long enough for some refractory nodes to recover and provide substrate for further self-enhancing
cycling excitation. Actually, the analysis of the simulated dynamics in its layered representation
for several network realizations shows that as soon as a
hole appears in the first layer, other holes rapidly appear in subsequent
layers, thus supporting the possibility of cycling excitations, possibly
numerous ones, and the sharp increase of the output signal. 
This mechanism to get re-entering excitation is quite similar to the
mechanism for achieving curling in spiral wave formation \cite{Geberth:2008p525,Liao:2011be,Garcia:2012ey}: There must
be a gap in the propagating front. Here either the excitation
propagation meets a refractory node, or it fails to excite all the
susceptible nodes that it encounters. It actually seems that any small perturbation of the sequential excitation of layers (observed in the low-$\kappa$ regime) is sufficient to trigger a full, self-sustained
response with nearly saturated output nodes excitations.

%Excitation
%fronts propagate, followed by a lagging front of refractory nodes
%preventing backward propagation, unless there is a `hole'
%in the excitation front, that is, a barrier that is not excited at
%once. A hole does not necessarily trigger a cycle; it may also enable
%longer paths, arriving later at the output node (note that excitations
%do not accumulate: at a given moment, the excited output node contributes
%by 1 to the output signal, whatever the number of excited neighbors
%triggering it). For instance, the hole is excited a step later by
%concurring excitations coming from other nodes of the same layer,
%or two steps later by concurring excitations coming from other nodes
%%of the next layer.
%A small perturbation of the sequential excitation of layers (observed
%in the low-$\kappa$ regime) is sufficient to trigger a full, self-sustained
%response with nearly saturated output nodes excitations

\begin{figure}
\includegraphics[width=0.95\columnwidth]{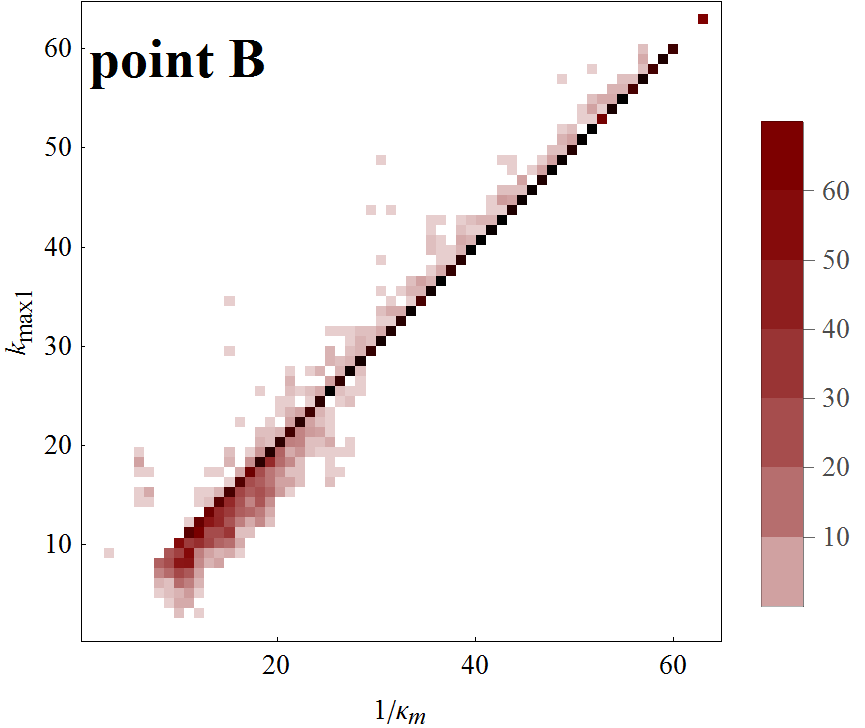}

\caption{A density histogram of the predicted degree for the onset of sustained
activity (transition point B) as a function of $1/\kappa_{m}$. The
data is aggregated over graphs having $N=80$ nodes and $M=100...2000$ edges.
The prediction is that $1/\kappa_m=k_{max,1} (the maximal degree in the first layer)$.
\label{fig:density-histogram}}
\end{figure}
\begin{figure}
\includegraphics[width=0.47\textwidth]{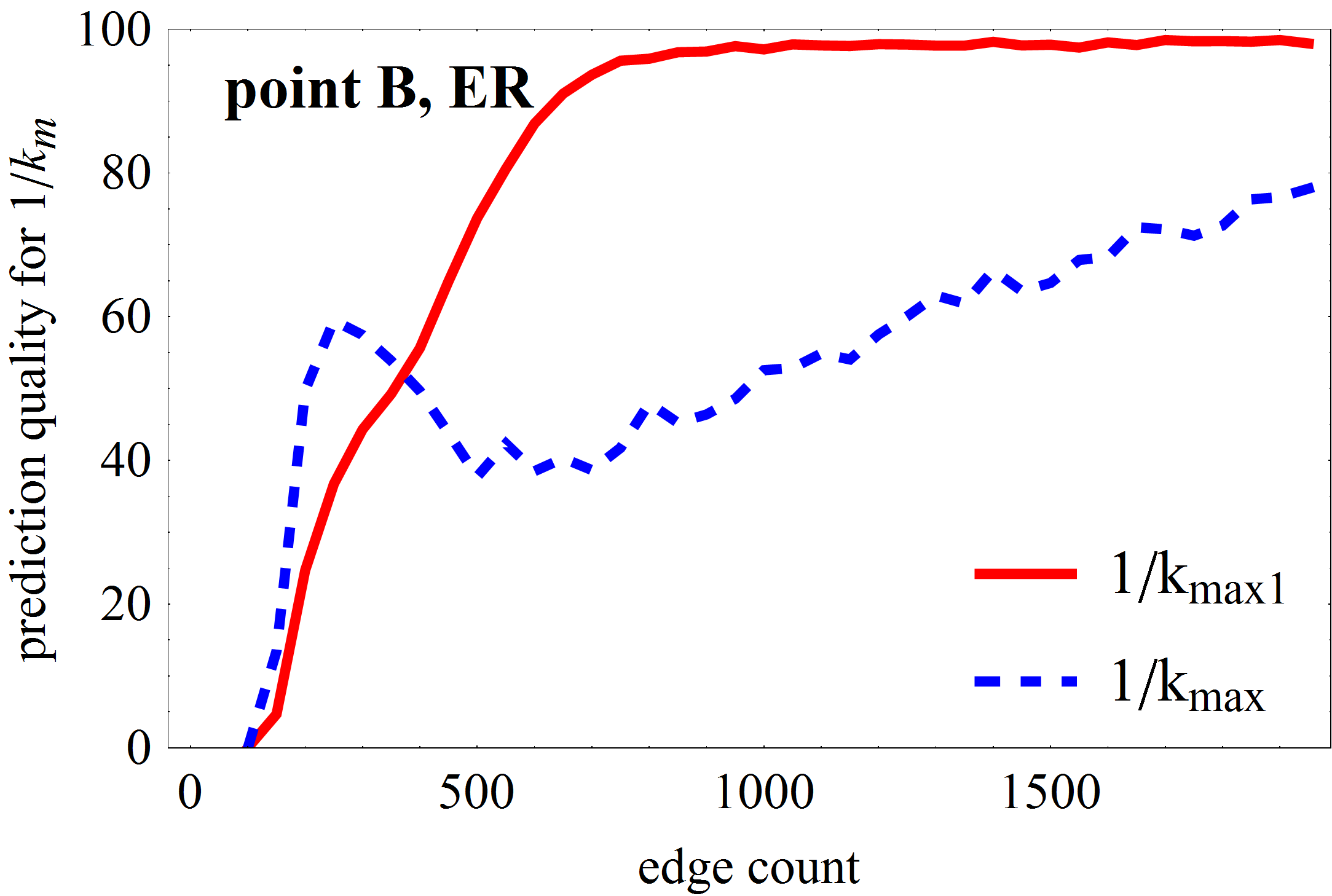}
\includegraphics[width=0.47\textwidth]{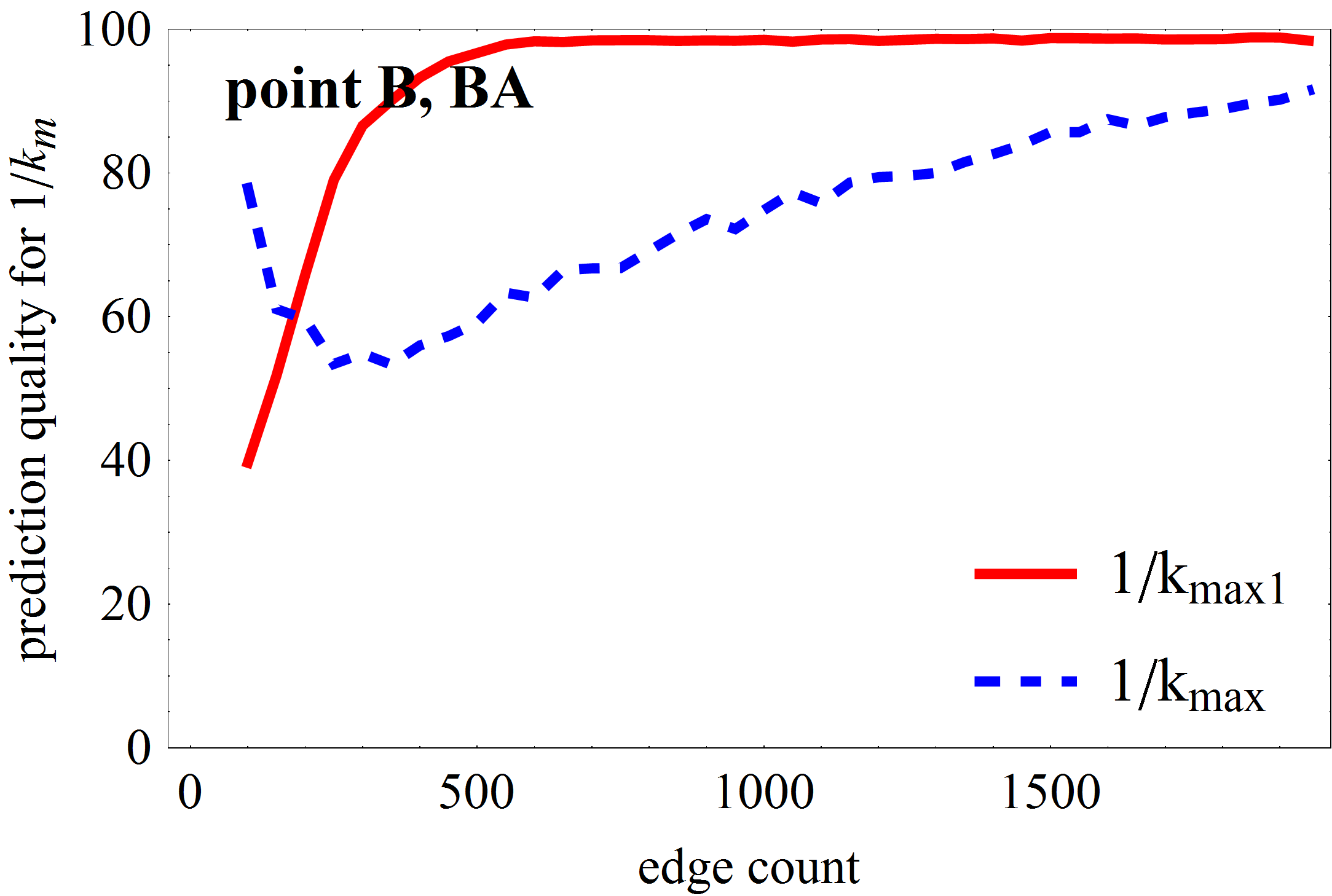}

\caption{Prediction quality for the onset  $1/\kappa_{m}$ of sustained activity.
The data are obtained  for graphs having $N=80$ nodes  and $M=100...2000$ edges. The predictions
are  $1/\kappa_{m}=k_{max,1}$ (i.e., the maximal degree in the first layer, red solid curve) and $1/\kappa_{m}=k_{max}$ (i.e., the maximal degree, blue dashed curve). The upper panel is for ER graphs, while the lower panel is for BA graphs. \label{fig:predQualOnset}}
\end{figure}

Denoting $k_{max}$ the maximal degree encountered
in the network, a rough estimate of the jump
location is $1/\kappa_{m}=k_{max}$.   In fact, the degree distribution being layer-biased, it is expected that with
a high probability the first hole appears in the first layer. Accordingly, 
another prediction is $1/\kappa_{m}=k_{max,1}$ where $k_{max,1}$  is  the maximal degree encountered in the first layer.

%Whether these predictions are valid  will depend on the exact pattern of
%cycles in the network realization, and hence varies from one network
%realization to another.
 The discrepancy with respect to our prediction of $\kappa_{m}$ is
expected to mostly originate in situations where the first `hole'
(node remaining susceptible while the excitation front propagates
downward the layers) is not located in the first layer. Additionally, an excitation hole does not necessarily trigger a cycle; it may also enable longer paths, arriving later at the output node (note that excitations
do not accumulate: at a given moment, the excited output node contributes
by 1 to the output signal, whatever the number of excited neighbors
triggering it). For instance, the hole is excited a step later by
concurring excitations coming from other nodes of the same layer,
or two steps later by concurring excitations coming from other nodes
of the next layer. This will also  contribute to  the discrepancy between our prediction and the actually observed value.

 We  compare our two predictions for $\kappa_{m}$, namely $1/k_{max}$
and $1/k_{max,1}$, for ER graphs (see Figure \ref{fig:density-histogram}).
% where $k_{max}$ is the maximaldegree observed in the whole network while $k_{max,1}$ is the maximal
%degree observed in the first layer.
Note that, as we have observed that the transition value  $\kappa_m$ does not depend on the value of $p$, we here
use $p=1$ (the deterministic case). 

For dense networks (right part of the curves in Figure \ref{fig:predQualOnset}), the prediction $1/k_{max,1}$ has a 100\% quality, meaning that a hole in the first layer is what
conditions, directly or indirectly, the onset of a significant amplification. This effect can be more directly observed in Figure \ref{fig:density-histogram}.
As the network gets denser, the number of layers  (the network maximal diameter) decreases, and the
size of the first layer increases. We might think that soon, the node
of maximal degree lies in the first layer. This is not the case, as
shown by the discrepancy between the quality curves for the prediction
$1/k_{max,1}$ and for the prediction $1/k_{max}$, which lies far
below. If the node of maximal degree was in the first layer, then
$k_{max,1}=k_{max}$ and the two curves would coincide. Hence, for
dense graphs the presence of a hole in the first layer is important for signal propagation, while the hub of maximal
degree is of no matter (although it would behave as a hole for smaller
$\kappa$).

On the contrary, for sparse graphs, the prediction quality for
$1/k_{max}$ outperforms the one based on $1/k_{max,1}$. 
When $\kappa$ is large enough for holes to appear in the first
layer and be involved in recurrent excitation (cycles), the signal
amplification is already working, due to a hole located in a deeper
layer and having a degree $k_{max}>k_{max,1}$.
%Signal amplification by recurrent
%(cycling) excitation relies (directly or indirectly) either on the
%node of maximal degree in sparse graphs (and appears as soon as this
%node is left as a hole by the first excitation front) or, in denser
%graphs, on the node of maximal degree in the first layer.%
%Let us focus on sparse graphs: w
What apparently matters most  for signal amplification by recurrent (cycling) excitation
is the
delayed excitation 
%\begin{comment}
%(via recurrent excitation? generating a cycle?)
%\end{comment}{}
 of a global hub. What apparently matters in dense graphs is the delayed
excitation of a hub in the first layer.
%, or at least, that the relativethreshold is such that also the hub in the first layer (and possibly
%in several other layers) is not excited from the single initial excitation.
At this point, it is difficult to say whether it is the presence of
a hole \emph{per se} which matters, or whether what matters are correlated
features (e.g. the presence of a sufficient number of holes, or some
more intricate feature of the available cycles).
These higher-order conditions for the signal amplification setting in at this threshold value of $\kappa $ are the reason for the low prediction quality in the case of sparse graphs.

In fact, understanding these curves and improving our predictions ask for a better understanding
of what happens after a `hole' as appeared in the
excitation front, and what are the requirements, in terms of either
cycle statistics or paths statistics or presence of other holes (i.e.
degeneracy of the degree $k_{max}$ or $k_{max,1}$), to get recurrent
activity.

\subsubsection{Height of the response curve (excitation level C)}

The activity level at point C is linked to the appearance of cycling excitation, feeding (directly or indirectly) into the output node, up to the maximum where
the output node is almost periodically excited, with the maximum average period $2+1/p$  for each value of the recovery probability $p$.
In this region, the situation is presumably a set of redundant cycles,
ensuring maximal excitation, so that on average an output node has
one excitation every $2+1/p$ steps, yielding a (trivial) level of
excitation equal to 
\begin{equation}
\frac{T}{2+1/p}
\label{eqTime}
\end{equation}
where $T$ is the length of the recording. 

We observe almost maximal excitation densities at the output nodes. This suggests that a set of redundant cycles compensates the stochasticity generated by the recovery probability $p$.

\begin{figure}
\includegraphics[width=0.96\columnwidth]{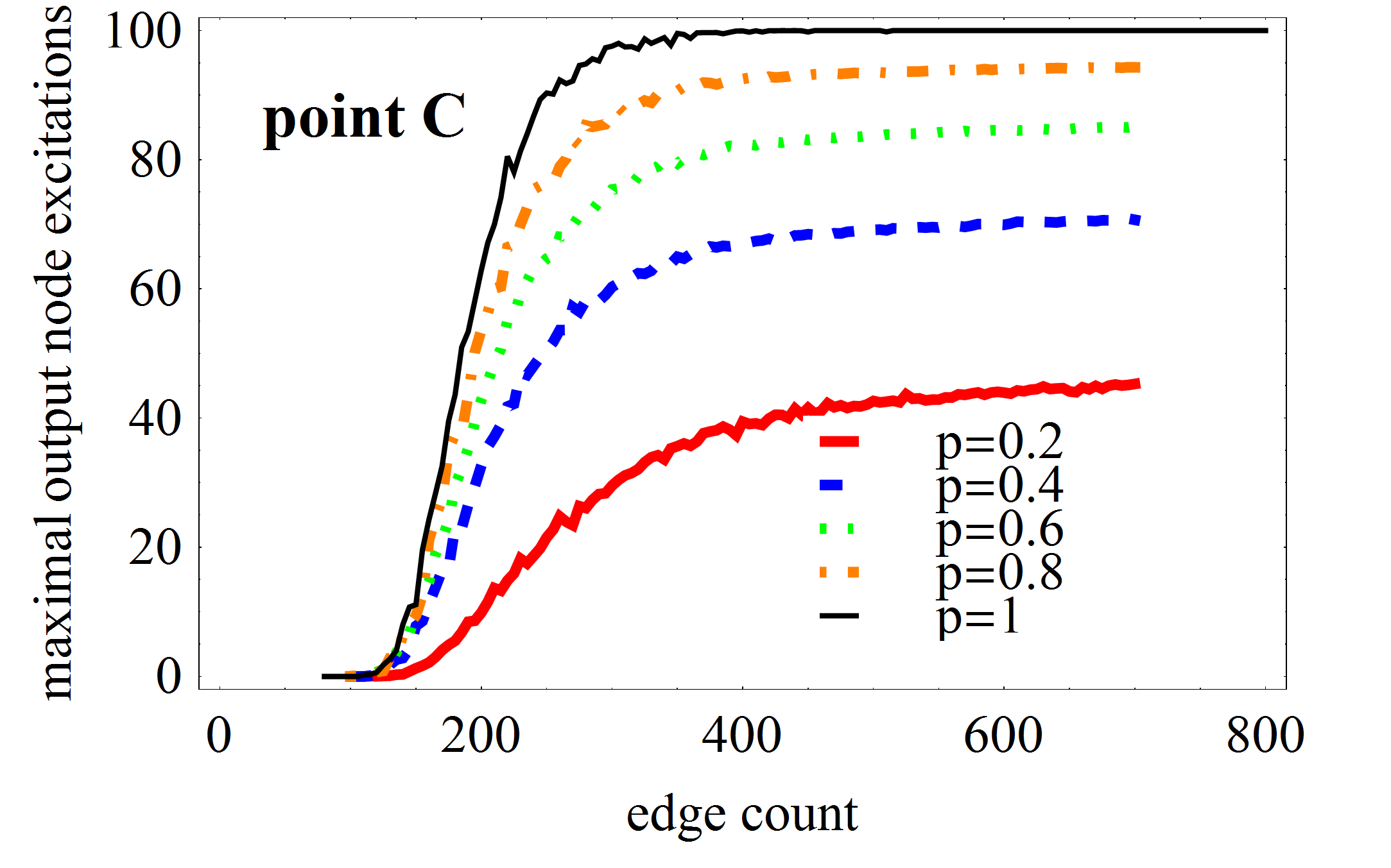}\caption{The maximum height of the output signal for different values of $p$.
The bottom level (output node) activity is normalized such that $100$ corresponds
to the maximal capacity of the output node at $p=1$. The number
of edges $M$ is varied, while the number of nodes is constant ($N$
= 80). For dense graphs, the sustained activity saturates to a value
dependent on $p$. The number of output excitations is aggregated
over $T=600$ steps, so that shorter transients do not matter.\label{fig:maxActVarPER}}
\end{figure}

\begin{figure}
\includegraphics[width=0.96\columnwidth]{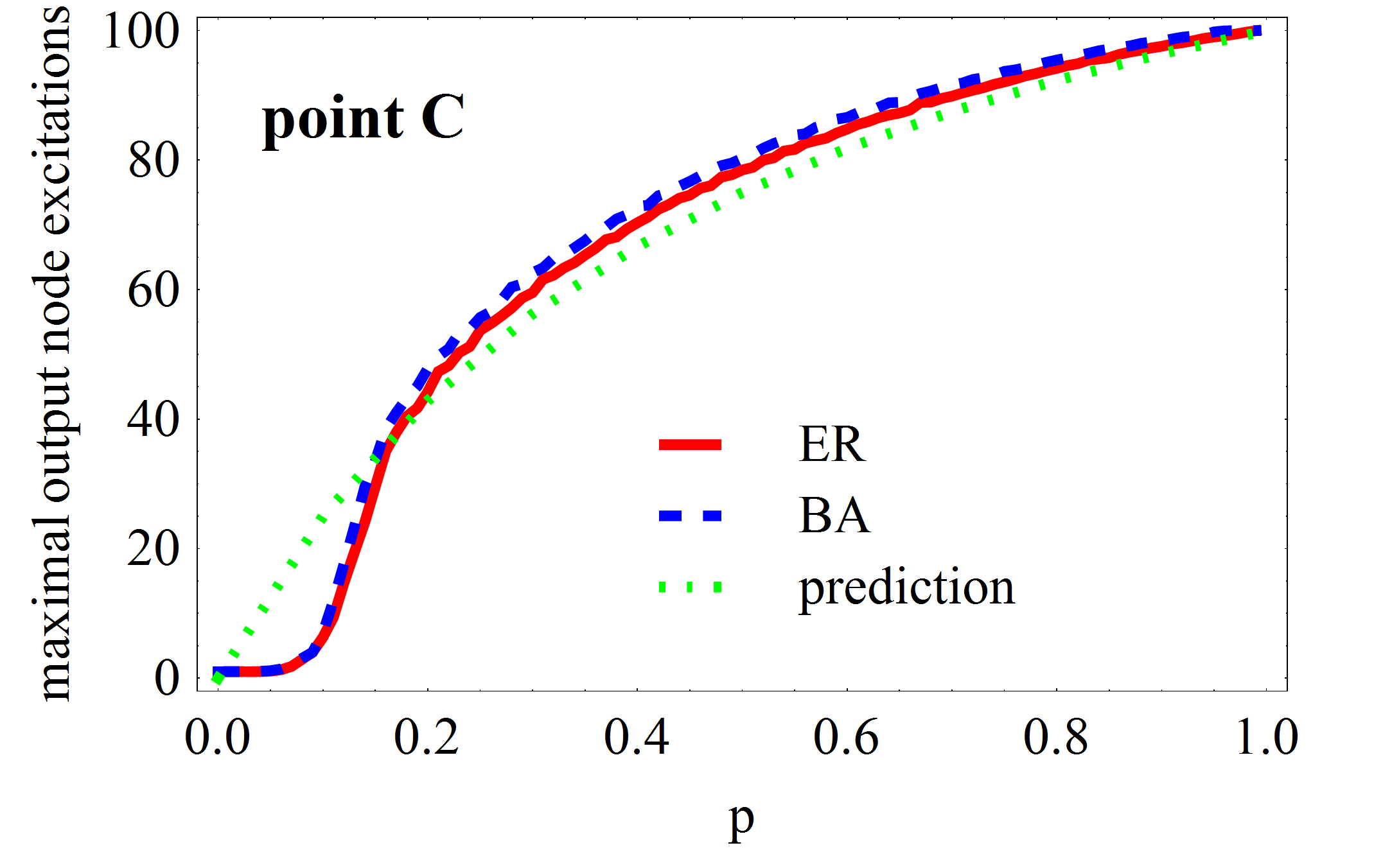}\caption{The scaling of the bottom level (output node) saturation activity over the recovery
probability $p$ for dense ER and BA graphs ($N=80$,$M=640$). The
output node activity $a$ is normalized such that $100$ corresponds to
its  maximal capacity, at $p=1$. The ER graphs (solid red curve) and the BA graphs (dashed blue curve) behave very similarly. Additionally
a prediction curve is included ($a=300/(2+1/p)$) in dashed green.} \label{maxActVarPSat}
\end{figure}

Figure~\ref{fig:maxActVarPER} shows the maximal height of the response
curve for various values of $p$ as a function of the edge density. As more
and more cycles are formed by the added edges, the output node excitations
quickly saturate at a value to a $p$-dependent level.
The scaling of the output node saturation activity as a function of the refraction
probability $p$ for dense graphs is shown in Figure~\ref{maxActVarPSat}.

The results are normalized so that its maximal capacity, for $p=1$, equals $100$, so that $a=300/(2+1/p)$. 
Generally, both the curves for the ER and the
BA graph fit the prediction well. This prediction,  merely equal to the output node capacity, constitutes an upper bound.

For Figure \ref{fig:maxActVarPER} we pick the maximum value of the output node excitation under variation of $\kappa $. In this way our numerical curve slightly overestimates the average maximum value predicted from Eq. (\ref{eqTime}). 

\begin{figure}
\includegraphics[width=0.96\columnwidth]{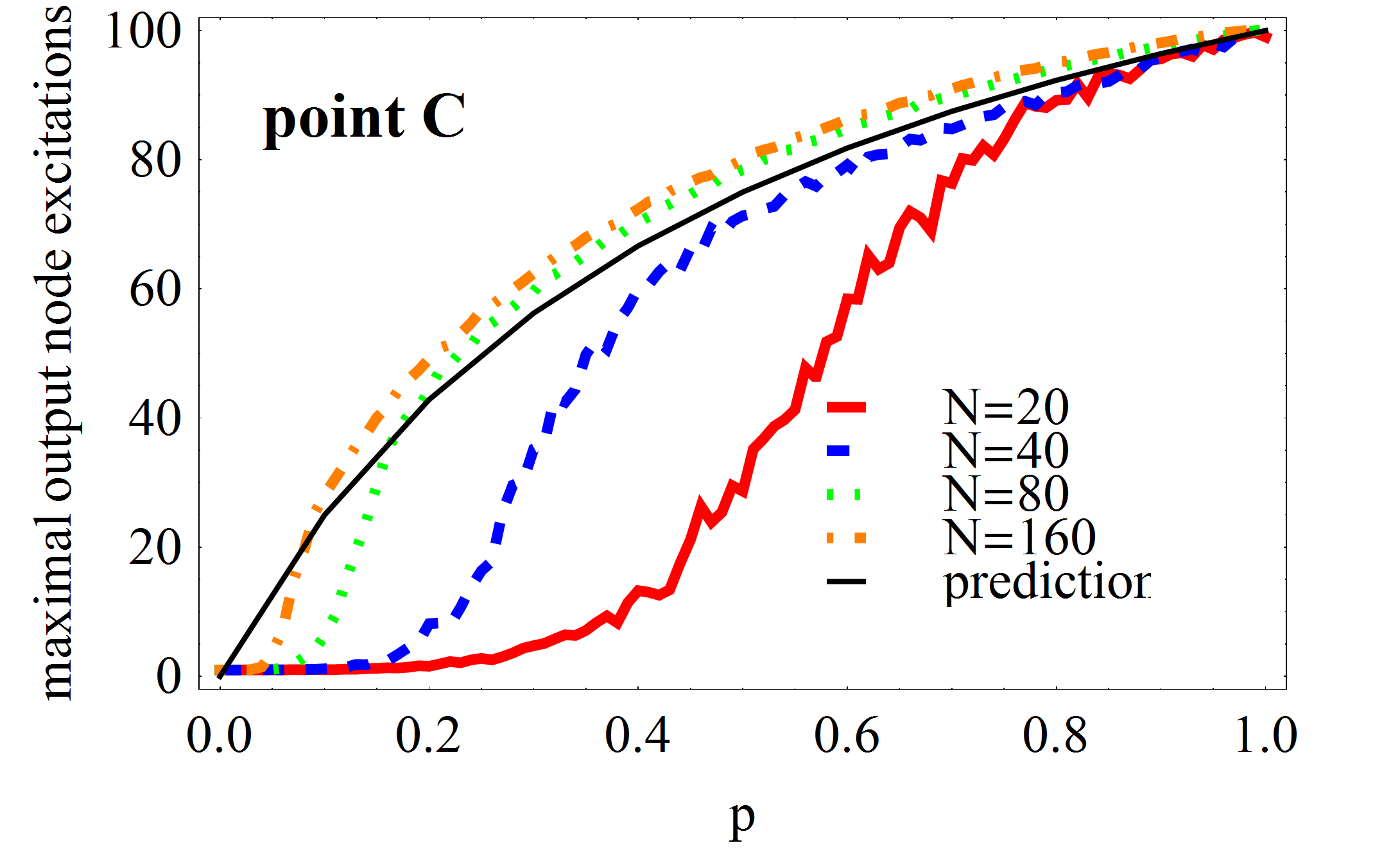}\caption{The scaling of the bottom level (output node) saturation activity over the refraction
probability $p$ for dense ER graphs of different sizes. The output node
activity $a$ is normalized such that $100$ corresponds to its
maximal capacity at $p=1$. Additionally a prediction
curve is included ($a=300/(2+1/p)$) as a thin black line. \label{maxActVarPSatMS}}
\end{figure}

The kink for small $p$ is due to an inability to sustain the activity
due to small graph size combined with many refractory nodes. This
finite size effect is further investigated in Figures \ref{fig:maxActVarPER} and \ref{maxActVarPSatMS},
confirming that it disappears for larger graphs.
This point emphasizes again the importance of studying small or medium-sized graphs, rather than just the asymptotic limit of infinite graphs, as real-world graphs across all domains of application (from biological to social and technological networks)  tend to be comparatively small (with numbers of nodes mostly in the hundreds). In small networks, the topological details like the arrangement of cycles and the barrier structure are of importance for qualitative features of the dynamics, while for infinite graphs these details can be expected to average out.

When $p=1$, the dynamics is deterministic and sustained activity
originating from robust pacemakers becomes possible, such as the triangle
$ESR$ or the square $ESSR$ (see also \cite{Garcia:2012ey}). This setup yields an output excitation increasing linearly with the duration of the observation $T$. For $p<1$, the excitation ultimately
vanishes in a finite network; however, for $p$ close enough to 1,
a long transient activity is observed, during which the accumulated
output signal increases with $T$. Practically the transient grows
exponentially with $p$ and is longer than any reasonable simulation
length, for example a network with $N=80$/$M=284$ reaches a transient
length of $10^{6}$ around $p=0.4$.

\section{Discussion and Conclusion}
\label{discussion}

The observed phenomena can be classified as `path-driven' (for the transition in A, $\kappa=\kappa_c$)
and `cycle-driven' (for the transition in B, $\kappa=\kappa_m$). 
Indeed, the transition
between sub-threshold (no propagation to the output nodes) and supra-threshold dynamics is due to the appearance
of the first barrier-free path.  On the contrary, the transition between simple signal propagation and signal amplification
is due to topological cycles and the possibility of cycling excitation
that occurs as soon as some nodes are not excited in the first stage
of signal propagation. Cycling excitation is involved at low $\kappa$,
explaining the amplification of the output signal once a hole as appeared
in the excitation front (currently in the first layer). 

The layer representation starting from a given node provides a node-centered
view that a given individual node may have of the network in which
it is embedded. This view is relevant in several instances, such as
the local probing of a network with no possibility to have an overall
and external view, e.g. probing the internet, propagation of signals
in neural networks, social networks in which an individual has only
a subjective view of the network to which s/he belongs, local control
of a logistic or engineered network in which only some localized nodes
can be acted upon.
At intermediary values of $\kappa$, the excitation dynamics is sensitive
to the hierarchical layer representation of the network. In this
sense, we have a process-induced layering, which could also happen
in real networks, of a few input nodes have been specifically selected
and evolved to match suitable topological features for the relevant
dynamics.

Our  simple numerical experiment and its interpretation provide a
reference case illustrating typical topological mechanisms that can
be at work in shaping the propagation and amplification of a signal
in an excitable network. Among the mechanisms we specially underline propagation due to a huge
path redundancy and
amplification due to cycling excitations. Our study enlightens the topological preconditions of spontaneous
activity, which is of relevance  to understand which
topological properties of a neural network enhance resting state activity \cite{Deco:2009p6486,Deco:2011p775}.
Moreover, these properties also form the precondition for the specific
reverberations that may serve as a dynamic representation of memory.
%Overall, our toy model relates to concrete issues of understanding
%resting state activity in brain networks {[}Deco et al. 2009{]}. and
%hints at some topological requirements for sustained activity. 
When $p<1$, cycle multiplicity seems essential to sustained activity, because each individual cycle will have 
a very limited activity. The values of $\kappa_{m}$ and $\kappa_{c}$
provide a way to calibrate different graphs when investigating for instance the influence of the architecture on the dynamic behavior. The difference $\kappa_{c}$-$\kappa_{m}$
can be taken as a unit for $\kappa$.

A vast amount of studies attempted to understand how network topology affects simple dynamical processes. Examples of such processes include synchronization \cite{Arenas:2006ba,deArruda:2013hd}, random walks \cite{Menezes2004FluctuationsinNetwork} and the propagation of excitations through networks \cite{MullerLinow:2006ex,Garcia:2012ey}. 

We explored how network topology determines the probability of dynamical events regulating the onset of persistent excitable dynamics (transition point A, Figure 3) in a graph, as well as the transition from propagating waves to sustained activity (transition point B, Figure 5) as a function of the relative excitation threshold. We use single excitations to probe the networks' dynamical capabilities as a minimal numerical experiment to gain insight into the mechanism underlying these two transitions. Our investigation thus sheds light on a situation of high interest to statistical physics: How do network details determine the propagation of excitations through a given network. We find that the excitation threshold selects certain topological constellations in the network, which serve as dynamical seeds initiating these transitions. 

Here, each graph has its own individual thresholds for the two transitions. Our mechanistic understanding of the dynamics is sufficient for predicting these two thresholds on the basis of topological information alone. This statement is validated by evaluating the prediction quality across a wide range of graphs.

Two result have been described in this paper: (1) For a specific graph, we can predict the
critical threshold values.
In spite of the similarities on the qualitative level, the response curves can look very different in the details (transition points, height) depending on the specific choice of the input node. 
This is due to the fact that, seen from one input node, a highly specific barrier structure on the paths towards the output node is encountered, as well as a specific arrangement of cycles along these paths. 
As we have demonstrated with the numerical experiments described in this paper, these topological details directly affect the response curve. 
(2) Our investigation draws the attention to a new network property: the barrier and cycle structures of networks, when hierarchized from specific input nodes. 
In evolved networks (like cortical area networks) this observation suggests the possibility of identifying input and output nodes via an optimized (or evolutionarily shaped) barrier and cycle structure along the interlinking paths. 

%The present study investigates very general parameters of sustainedactivity. 
In order to make the study more relevant for understanding  sustained
activity is real-world and particularly neural networks, it would naturally
be very interesting to expand the focus to (i) structured, non-random
networks (e.g., what would be effect of ring lattice \cite{Vishwanathan:2011it},
modular or hierarchical architectures), (ii) consider the different
dynamic patterns induced by specific stimulation of different input
nodes. In particular understand, to what extent do such specific stimulations
lead to reproducible patterns of activity.

However, our set of results can already be used in real cases or more complicated
numerical situations as a basis for delineating the contribution due
to these simple mechanisms and the contribution due to the involvement
of additional and more specific mechanisms. Unraveling the coupled
dynamical and topological origin of the different features of the
curve clearly shows the articulation between a regime dominated by
cycling excitation and a regime controlled by barriers along linear
(possibly redundant, as in an oriented mesh) paths.

\section*{Acknowledgement}
The authors are supported by DFG grants HU 937/7-1, HI 1286/5-1 and SFB 936/A1.

\newpage{}

%\bibliographystyle{spmpsci}
%\bibliography{thresholdBibV1}
%\input{ThresholdCurve-v16.bbl}

\end{document}